\documentstyle[12pt]{article}

\newcommand{\be}{\begin{equation}}
\newcommand{\ee}{\end{equation}}

\begin{document}

\baselineskip=18.0pt

\begin{center}

{\Large {\bf Self Avoiding Walks in Four Dimensions: Logarithmic Corrections}

\vspace{1.cm}}

{\bf Peter Grassberger, Rainer Hegger}

Physics Department, University of Wuppertal, D-42097 Wuppertal, FRG

and

{\bf Lothar Sch\"afer}

Physics Department, University of Essen, D-45117 Essen, FRG

\vspace{.8cm}

\today

\vspace{1.1cm}

{\bf Abstract} \end{center}

{\small \advance \baselineskip by -2pt

We present simulation results for long ($N\leq 4000$) self-avoiding
walks in four dimensions. We find definite indications of logarithmic
corrections, but the data are poorly described by the asymptotically
leading terms. Detailed comparisons are presented with renormalization
group flow equations derived in direct renormalization and with results
of a field theoretic calculation.
}

\eject

\section{Introduction}

Self avoiding walks (SAW's) are of practical importance since they
form a model for randomly coiled linear polymers. But even more
important for theoretical physics is that they represent in some sense
the simplest critical phenomenon. More precisely, they formally are
described by the $m\to 0$ limit of the $O(m)$ Landau--Ginzburg field
theory~\cite{dege}. Other members of this family are the Ising ($m=1$)
and the Heisenberg ($m=3$) models.

The most profound theoretical understanding of these critical phenomena
is obtained by the field theoretic renormalization group evaluated near
4 dimensions. Above $d=4$, all $O(m)$ models show mean field
behavior. Below $d=4$, the deviations of the critical exponents from
their mean field values are of order $\epsilon = 4-d$. This follows from
the fact that the renormalized coupling constant is of order $\epsilon$
in the infrared limit. The most extensively studied method for estimating
critical exponents (the ``$\epsilon$-expansion") involves resumming
perturbation expansions in $\epsilon$. Precisely at
$d=4$, the deviation from mean field behavior is given by logarithmic
corrections which can be predicted unambiguously by renormalization
theory.  Specifically, the leading
behavior in the limit of long chains is \cite{dege}
\be
   R^2_N \; \sim \;N\,[\log N]^\alpha\,,\quad \alpha = 1/4   \label{rr}
\ee
for the average squared end-to-end distance, and
\be
   C_N \; \sim \; \mu^N \,[\log N]^\beta\,,\quad \beta = 1/4  \label{cc}
\ee
for the number of distinct walks. These predictions are basic results.
In particular, they do not suffer from any ambiguities inherent in a
resummation of the $\epsilon$--expansion. It thus would
be extremely useful if one could verify them by independent means.
The most obvious candidates for alternative calculations
are exact enumerations of short walks and Monte Carlo
simulations. Both methods have been applied in the past.

Enumerations of chains with length up to $N=18$ have been used in
\cite{mck,gut,mcd} to verify eq.(2) ($R_N$ was not computed in these
papers), and recently chains were enumerated with $N$ up to $N=21$
\cite{mcd2}. From this the authors claim excellent agreement with
eq.(2): the power of the logarithm in the best fit is $\beta
=0.250\pm 0.005$.

Monte Carlo Simulations, on the other hand, have been used previously
for estimating $\alpha$ in \cite{havlin,rapa}. In addition, different
exponents related to $\alpha$ and $\beta$ were measured in
\cite{aragao,rens}. While refs.
\cite{havlin,aragao,rens} claim good agreement with eq.(1) (with  e.g.
$\alpha =0.25\pm 0.02$ in \cite{havlin}) and with eq.(2), serious
disagreement was found in \cite{rapa}: the best fit was obtained there
with $\alpha= 0.31$. Although no error bars were given in \cite{rapa},
the author obviously considered the value $\alpha = 1/4$ to be ruled
out. This seems a serious problem since the simulations of
ref.\cite{rapa} are by far more significant statistically than those
of \cite{havlin,aragao,rens},
both concerning the chain lengths ($N$ up to 2400) and the sample size.
Also, the method of analysis used in \cite{havlin} was justly criticized
in \cite{rapa} since it introduces an uncontrolled bias and does not
use the data optimally.

The methods used in \cite{aragao,rens} (in both papers, identical
methods were used) are not easily compared to the present one. In
different runs, these authors measured the average length
$\langle N_R\rangle_p$ of chains with fixed end-to-end distance $R$
in grand canonical ensembles at $p<p_c$ ($p$ is the fugacity),
and the probability $Q(p)$ that two chains starting at neighboring
sites never cross each other. We have not measured $Q(p)$ ourselves
(such measurements will be presented in \cite{hegger}), but we have
measured $\langle N_R\rangle_p$. From this we shall argue in sec.5
that the analyses of \cite{aragao,rens} have large systematic errors,
and that indeed deviations from mean field behavior is much larger
than claimed there.

In view of this situation (and since simulations of theta polymers in
3 dimensions \cite{theta} also gave discrepancies with the logarithmic
corrections expected there \cite{dege}) we decided to perform
simulations with much higher accuracy than those done previously.

We find that the numerical results of \cite{rapa} are correct (though
we do not agree with the conclusions drawn from them).  In contrast,
it seems that the findings of \cite{havlin,aragao,rens} were not
completely uneffected
by wishful thinking. This means that the leading logarithmic terms of
(\ref{rr}) and (\ref{cc}) are not sufficient to describe the behavior
at any chain lengths which can be simulated in the foreseeable future
(unless we assume, as was done in \cite{rapa}, that the leading terms
show a different power of $\log N$).

The first corrections to the asymptotic laws (\ref{rr}) and (\ref{cc})
have been calculated by Duplantier \cite{dupl}, but it turns out that
even the corrected expressions are not fully consistent with the
numerical data. These expressions involve a non-universal parameter
(an integration constant from integrating the renormalization group
flow) which should be the same for $R^2_N$ and $C_N$. But for a good
fit two different values have to be chosen for $R^2_N$ and $C_N$, which
is an internally inconsistent procedure. However, from our data we can
extract logarithmic derivatives $\partial\ln R^2_N/\partial\ln N$,
$\partial\ln C_N/\partial\ln N$, which may be compared directly to
renormalization group flow equations of an appropriate direct
renormalization scheme. Such a comparison does not involve unknown
fit parameters. It thus is very pleasing that the flow equations
consistently can be put into a form which compares very well to our
Monte Carlo data.

Stimulated by that success we use the original data to test another
theoretical approach. As has been stressed previously \cite{schaef,kraus},
the renormalization approach can be viewed as proceeding in two
essentially independent steps. We first determine a mapping from the
physical model to its renormalized counterpart, and in a second step
we use renormalized perturbation theory to determine the observables,
working completely within the renormalized model. It has been
suggested that the mapping must be determined most precisely, but in
the second step we may be content with low order perturbation theory. We
find that this method allows for a consistent determination of the
nonuniversal parameters and yields a good quantitative fit to our data.

The paper is organized as follows: In the next section we present
our Monte Carlo algorithm and the resulting data. Comparison with
renormalization group flow equations derived in the spirit of direct
renormalization is given in sec.3, while comparison with field
theoretic renormalization is presented in sec.4. We conclude with a
discussion of our results in sec.5.

\section{Simulations}

Aiming at estimates for both $R^2_N$ and $C_N$, we decided not to use
the pivot algorithm \cite{pivot}, though it should be the most efficient
algorithm for estimating $R_N$ in the limit $N\to\infty$. Instead, we
essentially used a recursive and randomized implementation of the old
enrichment method \cite{wall}. This method is an improvement over the
incomplete enumeration or `recursive sampling' method used in
\cite{rr,gra}, and was applied successfully in \cite{theta,gra-heg}
to polymer adsorption on surfaces, to theta polymers and to off-lattice
polymers interacting via Lennard-Jones potentials.

The basic structural element of incomplete enumeration is a subroutine
STEP({\bf x}) which marks the site {\bf x}
as occupied and calls itself at all neighboring sites
${\mathbf x}\pm {\mathbf e}_i$, provided these sites are still free and
provided $N<N_{max}$. Before leaving the subroutine, the
site {\bf x} is marked as free again. If each free neighboring site were
visited with probability 1, this would give exact enumeration. In order
to obtain a grand canonical distribution of walks where
\be
   n_N = {\mathrm const} \;C_N p^N
\ee
is the average number of $N$-step walks, one has to use a random number
generator so that only a fraction $p$ of all free neighbors are visited.

The basis of the present improvement is the observation that it is
sufficient to visit only one of the neighbors (provided it is free;
otherwise the subroutine is left immediately), but in average $({\cal N}-1)p$
times. Here ${\cal N}$ is the coordination number of the lattice, and we
have assumed that we do not attempt any back steps as the corresponding
site would not be free anyhow. The average chain length diverges when
$p\to p_c = 1/\mu$. Since $({\cal N}-1)p_c > 1$ for all lattices, this
means that we make $>1$ attempts to continue each successful path if
$p\approx p_c$. Thus we choose a random neighbor (different from that
we had come from), call STEP at this neighbor, and after having returned
we call STEP again at the same neighbor with probability $({\cal N}-1)
p_c-1$. The main difference to the algorithm of \cite{bs}, e.g., is
that we always try at least once to continue the walk.

Like the incomplete enumeration method of \cite{rr,gra} and like the
method of \cite{bs}, the present method corresponds essentially to
a random walk in the chain length with reflecting boundary conditions at
$N=0$ and $N=N_{max}$. At $p_c$, it takes thus roughly $N^2/D$ steps to
obtain one
statistically independent SAW of length $N$. Here $D$ is a diffusion
constant which is of order unity in the algorithms of \cite{rr,bs,gra}.
The main advantage of the present algorithm (apart from slightly shorter
programs) is that here
\be
   D\sim 1/\delta\,,\quad \delta =(({\cal N}-1)p_c -1).     \label{D}
\ee
Since $\delta \approx 0.033$ for the simple hypercubic lattice in $d=4$,
this gives a very large diffusion coefficient, and the method is more
than one order of magnitude faster than those of \cite{rr,bs,gra}.
Eq.(\ref{D}) can be
understood as follows: in the present algorithm, a step back in $N$ is
only needed when the walker hits an occupied site. The chance for this
is $\approx \delta$. In simulations near $p_c$, this means that in
average $1/\delta$ forward steps are made until one back jump reduces
$N$ by $1/\delta$. In each of \cite{rr,bs,gra}, in
contrast, the probability to make a back step is of order 1 (it is
$(1-p)^\mu$ for \cite{rr,gra}, and $1/(1+p{\cal N})$ for \cite{bs}).

We have simulated SAW's of length up to $N_{max}=4000$. Since it
is not possible to store the sites occupied by such long walks in a
simple bit map, we used a hashing procedure similar to (but somewhat
simpler than) that used in \cite{pivot}. We made only
simulations very close to $p_c$, where $n_N$ is roughly independent of
$N$. As explained in \cite{wall,gra}, this should be most efficient. Our
total sample corresponded to $n_N \approx 10^8$. This is also roughly
the number of SAW's of maximal length $N_{max}$, but all these walks are
of course not independent. Often it is stated that this correlation
between the walks is the main drawback of the enrichment method. In our
version, it is not a big problem because of the ease with which walks are
generated. The number of independent walks is given by the number of
instances where the algorithm has reached $N=N_{max}$ between two returns
to the main routine (corresponding to $N=0$). Our sample contained
$\approx 1.2\times 10^6$ such independent walks. The total CPU time was
ca. 800 h on a cluster of DEC ALPHA workstations.

Our results for $R_N^2 / N$ are shown in fig.1. In order to compare with
eq.(1), we plotted its logarithm against $\ln (\ln N)$. To demonstrate
the importance of non-leading terms in logarithmic expressions, we show
two slightly different quantities. For the solid line, $N$ is defined as
the number of {\it bonds}, while $N$ is the number of {\it sites} for the
dotted line. These two definitions differ by one unit and are clearly
equivalent for $N\to\infty$. Nevertheless, we see that the difference is
still important for $N=300$! We show also the
results of \cite{rapa} which have larger error bars but are otherwise in
perfect agreement. From eq.(1) we expect our data to fall onto a straight
line with slope 1/4. This is definitely not seen. Instead, the best
straight line to the data (where $N$ is interpreted as the number of
bonds) has slope $0.311\pm 0.003$. It would indeed give a perfectly
acceptable fit.

Our data for $C_N$ are shown in fig.2. More precisely, we there plotted
$C_N p^N$ against $\ln N$ for three different values of $p$. We also show
the exact enumeration data of \cite{mcd} with which we are in perfect
agreement. But again we see hardly any sign of the predicted asymptotic
behavior. We do not want to discuss in detail why several authors
\cite{mcd,gut,mck}
were able to extract the correct asymptotic behavior from enumeration
data (for noiseless data there exist very sophisticated methods to extract
singularities), but obviously non-leading contributions are very large.

\section{Higher Order Terms and Renormalization \newline Group Flow}

Fortunately, the leading corrections to eqs.(\ref{rr},\ref{cc}) have
been calculated in \cite{dupl}
\be
  \alpha_N \equiv R_N^2 / N = r [\ln(N/a)]^{1/4} \left[ 1 - {17 \ln(4
     \ln(N/a)) +31 \over 64 \ln(N/a) } + \ldots\right]   \label{rr-dupl}
\ee

\be
  C_N /\mu^N = c [\ln(N/a)]^{1/4} \left[ 1 - {17 \ln(4 \ln(N/a))
     -3\over 64 \ln(N/a) } + \ldots\right] .            \label{cc-dupl}
\ee
The constant $a$ has to be treated here as a free parameter. We thus have
two parameters $(r,a)$ to fit $R_N^2 / N$, and two more ($\mu,c$)
if we also want to fit $C_N$.

The best fit to eq.(\ref{rr-dupl}) is obtained with $r=1.331, a = 0.1237$.
It is included in fig.1 (dashed line). Over the range of interest it
practically is undistinguishable from a straight line with slope 0.311.

For $C_N$ it is even more obvious that the leading term $[\ln(N/a)]^{1/4}$
would give a very poor fit. A fit with eq.(\ref{cc-dupl}), using $a$
again as a free parameter, gives $c=1.05, a = 0.055$. This fit is shown
as dotted curve in fig.2. Notice that the values of $a$ extracted from
$\alpha_N$ and from $C_N$ differ considerably. No acceptable fit is
found with a common value of $a$.

Neglecting this problem for the moment, our data suggest that the critical
value of $p$ is $ p_c = 0.147622 \pm 0.000001$.
This is to be compared to $p_c = 0.147625\pm 0.000002$ as obtained in
\cite{mcd2}. The values accepted in \cite{aragao,rens} were substantially
larger, $0.1490\pm 0.0003$ and $0.1493\pm 0.0007$, which partly explains
why smaller logarithmic corrections were found in these papers (see
sec.5).

We have thus been able to produce individual good fits to $R_N^2 / N$
and to $C_N p^N$, but the non-leading corrections are very important
(in the latter case masking completely the leading behavior), and the
fit parameters are not mutually consistent. This can also be seen
by plotting the ratio $(R_N^2 / N)/(C_N p^N)$. Here the dominant terms
cancel, and we obtain
\be
   {\alpha_N \over C_N p_c^N} = const \, [ 1 - {17 \over 32 \ln(N/a) }
            + \ldots]            \label{rc-supl}
\ee
{}From fig.3 we see that this ratio is indeed a roughly linear function
of $1/\ln(N/a)$, provided we take $a \approx 0.1$. The slope of this
function has the right sign but is roughly twice as large as the value
predicted by eq.(\ref{rc-supl}).
To summarize, even including the leading correction terms we do not
find a fully satisfactory explanation of the data.

To proceed we note that eqs. (\ref{rr-dupl},\ref{cc-dupl}) are derived by
integrating the renormalization group flow equations, keeping only
terms up to one loop order. It therefore is of interest to take a step
back and compare the data against the more basic flow equations. We
first reconstruct the flow equations from results given in the literature.

Following \cite{cloiz}, we start from dimensionally regularized
perturbation theory in $d=4-\epsilon$ dimensions. Denoting by $b$ the
bare coupling constant and by
\be
   z = {bN^{\epsilon/2}\over (2\pi)^{d/2} }
\ee
a dimensionless coupling strength, we have up to second order in $z$
\cite{cloiz}
\be
   \alpha_N = 1+z({2\over\epsilon}-1)+z^2(-{6\over\epsilon^2}
          +{11\over 2\epsilon}) + {\cal O}(z^3)  \label{first.alphan}\;,
\ee
\be
    C_N/\mu^N = 1+z({1\over\epsilon}+{1\over2})-z^2({7\over 2\epsilon^2}
          + {4\over \epsilon}) + {\cal O}(z^3)  \label{first.cn}\;.
\ee
Obviously these expansions are singular at $d\to 4$. We thus introduce
a renormalized coupling constant, which in the minimal subtraction scheme
of \cite{dupl2} reads
\be
   z_R = z-{8\over\epsilon}z^2 + ({64\over\epsilon^2}+{17\over\epsilon})z^3
          + {\cal O}(z^4)\;.
\ee
{}From this we get the Wilson function~\cite{dupl}
\be
  W[z_R,\epsilon] = N{\partial\over\partial N}z_R|_{b,\epsilon} =
       {1\over2}\epsilon z_R -4z_R^2 + 17 z_R^3 + \ldots\;,     \label{w}
\ee
in which we can take the limit $\epsilon\to0$ without encountering any
problems:
\be
  W[z_R] = -4z_R^2 + 17 z_R^3 + {\cal O}(z_R^4)\;.           \label{wz}
\ee
While $\alpha_N$ or $C_N/\mu^N$ expressed in terms of $z_R$ still are
singular, their derivatives with respect to $\ln N$ are known to be
finite. Indeed these derivatives yield flow equations governing the
renormalization of the chain length and of the partition function. Defining
\be
   \sigma_0[z_R,\epsilon] = N{\partial \ln\alpha_N \over\partial \ln N}
        \Big|_{b,\epsilon}        \label{sigma0}
\ee
and
\be
   \sigma_1[z_R,\epsilon] = N{\partial\ln C_N \over\partial \ln N}
        \Big|_{b,\epsilon} \;,      \label{sigma1}
\ee
we have \cite{dupl}
\be
   \sigma_0[z_R] = \lim_{\epsilon\to0}\sigma_0[z_R,\epsilon] = z_R+
       {7\over2}z_R^2+\ldots \; \label{sigma_0} \;,
\ee
\be
    \sigma_1[z_R] = \lim_{\epsilon\to0}\sigma_1[z_R,\epsilon] =
       z_R -5z_R^2 +\ldots \;.                                \label{sigma}
\ee

Equations (\ref{rr-dupl}) and (\ref{cc-dupl}) were obtained by first
integrating eq.(\ref{w}) at $\epsilon=0$, yielding $z_R$ as a function
of $N$. This was inserted into eqs. (\ref{sigma_0}) resp. (\ref{sigma}),
and eqs.(\ref{sigma0}) resp. (\ref{sigma1}) were integrated
again. During these manipulations, only the leading terms were kept,
since higher order terms are not completely known anyhow.

Avoiding these integrations we now directly compare our data to eqs.
(\ref{w})--(\ref{sigma}). We first calculate the derivatives
(\ref{sigma0}), (\ref{sigma1}) as functions of $N$.  Of course, we
have to replace derivatives with respect to $N$ by finite differences
(we use $\Delta \ln N = \ln2$ for first derivatives, and
$\ln4$ for second), but this should not present much problems in view
of the slow variations of all functions involved.

We thus show in fig.4
\be
   \sigma_0(N) \equiv {1\over\ln 2}\ln{R^2_{2N}\over 2R^2_N} \;, \label{sig0}
\ee
while
\be
   \sigma_1(N,p) \equiv {1\over\ln 2}\ln{p^N C_{2N}\over C_N}  \label{sig1}
\ee

is plotted in fig.5 for the same three values of $p$ as before. From
these plots, we should be able to obtain the same function $z_R(N)$ by
inverting eqs.(\ref{sigma_0}),(\ref{sigma}). A quick test shows that
this is not so easy. The problem is that $z_R(N)$ turns out to be not
very small (even for the largest values of $N$), and the Taylor
expansions (\ref{sigma_0}),(\ref{sigma}) obviously are poorly
convergent. This is particularly true for $\sigma_1$, for which
truncation in eq.(\ref{sigma}) after the quadratic term yields
$\sigma_1\le 0.05$, in contradiction to the data for $N<30$.

A trick which helps -- though its justification is far from obvious --
is to change the expansions in eqs.(\ref{sigma0}),(\ref{sigma}) into
Pad\'e approximants,
\be
   \sigma_0[z_R] = {z_R\over 1-7z_R/2}\;,\qquad
     \sigma_1[z_R] = {z_R\over 1+5z_R} \;.   \label{sigmap}
\ee
These can be inverted to give
\be
   z_R = z_R^{(\alpha)}=z_R^{(C)}
\ee
with
\be
   z_R^{(\alpha)}={\sigma_0\over 1+7\sigma_0/2}\;,\qquad
        z_R^{(C)}={\sigma_1\over 1-5\sigma_1}\; .   \label{zpade}
\ee

In fig.6 we have plotted $z_R^{(\alpha)}$ and $z_R^{(C)}$ as obtained
from the finite-difference approximations. We see still some
disagreement for small $N$ which may be due to higher order terms
in $\sigma_0[z_R], \sigma_1[z_R]$ and/or to $1/N$ corrections.
But for large $N$ we find very acceptable agreement, provided
we take $p=1/\mu=1.476223$.

The final test of the theory consists in checking whether these estimates
of $z_R$ satisfy the differential equation $N\,dz_R/dN = W[z_R]$ or,
rather, its finite-difference approximation. Again we find problems due
to the slow convergence of the Taylor expansion for $W[z_R]$, and again
we take recourse to a Pad\'e approximant,
\be
  W[z_R] = -{4z_R^2\over 1 + 17 z_R/4}  \;.           \label{wzp}
\ee
In fig.7a we have plotted this form of $W$ with argument $z_R^{(\alpha)}$
against $N$, together with
\be
   {\Delta z_R^{(\alpha)}(N)\over \Delta \ln N} = {z_R^{(\alpha)}(2N)
         -z_R^{(\alpha)}(N/2)\over \ln 4} \;.
\ee
We see very nice agreement, in particular for large values $N$ where it
is most significant. For small $N$ agreement could indeed be improved
by adding a constant of ${\cal O}(1)$ to $N$ so that $\alpha_N$ is
replaced by $R_N^2/(N+0.16)$, see fig.7b. Such a $1/N$ correction would
also improve the agreement in fig.6 at intermediate values of $N$.

We thus conclude that our data are fully consistent with
the renormalization group flow predicted theoretically, though there
is some ambiguity related to
the use of Pad\'e approximants and -- to a much lesser degree -- to the
treatment of $1/N$ corrections. We want to stress that this analysis
is particularly interesting since, except for the size of the
nonuniversal $1/N$ corrections, it does involve no fit parameter! Note
further, that this analysis uses only the information underlying also
eqs.(\ref{rr-dupl}) or (\ref{cc-dupl}). This suggests that solving the
renormalization group equations to one loop order and substituting the
result into expressions for higher order corrections we lose meaningful
information which still can be identified on the level of the flow
equations.

\section{Field Theoretic Renormalization}

As mentioned in the introduction, the renormalization program may be
separated into two steps: We first map the physical `bare' model on a
renormalized model which exhibits the scale invariance of long chains.
We then calculate $R^2_N$, $C_N$ perturbatively within the renormalized
model.

The bare model depends on the coupling constant $b$, the chain length
$N$, and a microscopic length scale $l$ which is of the order of the
lattice spacing in the computer experiments. In the renormalized
theory these parameters are replaced by the renormalized coupling $u$,
the renormalized chain length $N_R$, and a renormalized length scale
$l_R$. The mapping takes the general form
\be
   b=\mbox{const}\left(\frac{l}{l_R}\right)^\epsilon Z_u(u,l/l_R)u
                                                \label{sch.b}
\ee
\be
   N=\left(\frac{l_R}{l}\right)^2 Z_n(u,l/l_R) N_R\, ,
                                                \label{sch.N}
\ee
where the renormalization factors are normalized according to
\be
   Z_u(0,1)=Z_n(0,1)=1\,.                       \label{sch.zu}
\ee

The observables of interest here can be expressed as
\be
   R^2_N= 2d l_R^2 N_R A_R(u,N_R)   \label{sch.alpha}
\ee
\be
   \frac{C_N}{\mu^N}=\mbox{const}\frac{Z(u,l/l_R)}{Z_n(u,l/l_R)}A_C(u,N_R)
                                      \label{sch.cc}\,,
\ee
where $Z(\ldots)$ denotes another renormalization factor and the
amplitudes $A_R$, $A_C$ are to be calculated by renormalized
perturbation theory. In general the renormalization factors have to be
chosen to absorb the leading dependence on the microscopic scale $l$: for
fixed $l_R$, all observables have to become independent of $l$ up to
corrections of order $l^2/l_R^2N_R \sim 1/N$. Using the scheme of
dimensional regularization, defined by taking the limit $l\to0$ for $d<4$,
factors such as to make the renormalized theory finite in four dimensions.

The length scale $l_R$ is a free parameter of the renormalized theory.
Under a change of $l_R$ the other parameters change according to the
flow equations
\be
  \frac{\partial u}{\partial\ln l_R}=\epsilon u-\tilde{\beta}(u)
                            \label{sch.dudlnlr}
\ee
\be
  \frac{\partial\ln Z_n}{\partial\ln l_R}=\zeta_n(u)
                            \label{sch.dlnzndlnlr}
\ee
\be
  \frac{\partial\ln Z}{\partial\ln l_R}=\zeta(u)\,,
                            \label{schl.dlnzdlnlr}
\ee
where all derivatives have to be taken at fixed $b,l$ and $\epsilon$.
Based on Pad\'e--Borel summation of higher order perturbation theory, a
parametrization of the flow functions has been given
in~\cite{schloms}:
\be
  \zeta_n(u)=-u+\frac{5}{4}u^2-a_1u^3+a_2u^4   \label{sch.zetan}
\ee
\be
  \zeta(u)=-\frac{u^2}{4}+a_3u^3  \label{sch.zeta}
\ee
\be
  \tilde{\beta}(u)=4u^2\frac{1+a_4u}{1+a_5u}  \label{sch.beta}
\ee
with the constants $a_1=3.6328$, $a_2=3.8953$, $a_3=0.04395$,
$a_4=1.555$ and $a_5=3.5962$.

(Our notation differs somewhat from that of~\cite{schloms}: $u=8u^
{\mbox{\cite{schloms}}}$, $\zeta_n=-\zeta_r^{\mbox{\cite{schloms}}}$,
$\zeta=\zeta_\phi^{\mbox{\cite{schloms}}}$. We further note that recently
some mistake has been found~\cite{kleinert} in the five loop contributions
to the renormalization factors. However, due to the remarkable stability
of the Pad\'e--Borel results this is not expected to yield serious
changes in the parametrization of (\ref{sch.zetan})--(\ref{sch.beta}).)

Starting from arbitrary initial conditions set at $l_R=l$ we now
integrate the flow equations to find functions $u(l_R)$ e.t.c.\ in
analytic form. We then fix the final value of $l_R$ by the condition
\be
N_R=1 \label{sch.nr}\,,
\ee
which implies that $l_R$ is of order $R_N$. There result the expressions
\be
   \alpha_N = l_0^2 {e^{1.951u-1.422u^2+0.734u^3}\over
             u^{1/4}(1+1.555u)^{1.369}} A_R(u,1) \label{gr.alphan}\;,
\ee
\be
   C_N/\mu^N = c_0\; {e^{2.101u-1.434u^2+0.734u^3}\over
             u^{1/4}(1+1.555u)^{1.425}}A_C(u,1) \label{gr.cn}\;.
\ee

The renormalized coupling constant is implicitly determined by the
equation
\be
   N = n_0 {(1+1.555u)^{2.349}\over u^{0.730}} e^{{1\over 2u} -
          (1.951u-1.422u^2+0.734u^3)} \label{gr.n}\;.
\ee
Here $l_0$, $c_0$, $n_0$ are nonuniversal fit parameters absorbing the
initial conditions in the integration of the flow equations.

So far we have been concerned with the renormalization group mapping.
Calculation of the amplitudes is a straightforward exercise in
renormalized perturbation theory. It yields
\be
   A_R(u,1)=1-\frac{u}{2}+O(u^2) \label{sch.ar}
\ee
\be
   A_C(u,1)=1+\frac{u}{2}+O(u^2) \label{sch.ac}
\ee

Eqs.(\ref{gr.alphan})--(\ref{sch.ac}) constitute our final result.
{}From the derivation it should be clear that the calculation of the
amplitudes is a problem well separated from the determination of the
renormalization group mapping. Indeed, within the minimal subtraction
scheme it is the singular terms in expressions like
eq.(\ref{first.alphan}),(\ref{first.cn}) which determine the mapping,
whereas the amplitudes result from regular terms.

We now compare to the Monte Carlo data. Fitting
$\alpha_N$ in the range $300\leq N\leq 4000$ we find $l_0 = 1.0436$
and $n_0 = 0.275$ (see fig.8). The same value for $n_0$ together with
$c_0 = 0.864$ would also give an acceptable fit to $C_N$. But the best
fit to the latter (see fig.9a) is obtained with $n_0=0.316,\;
c_0=0.870$, and
\be
   p_c=0.1476223\pm0.0000005\;.
\ee
The latter value for $n_0$ gives also an excellent fit to $\alpha_N$,
provided $N$ is replaced by $N+n_1$ in the definition of $\alpha_N$,
with $n_1=0.56$ (see fig.9b). In the latter fit, $l_0=1.0466$. A priori,
such a change would be well within the uncertainty intrinsic in the
definition of $N$. But fig.7 and analogous plots of the dependence of
$u$ on $l_R$ in the present scheme suggest that such a value of
$n_1$ is somewhat large. We must note, however, that introducing $n_1$
we can account for $1/N$ corrections only in a very crude way
\cite{krueger}. We thus find agreement with the field theoretic
prediction, but there remains some room for further improvement in the
region of short chains.

It is of some interest to note the range of the renormalized coupling
resulting from our fit:  $0.067<u<0.15$ for $N$ in the range
$4000>N>100$. Thus even for $N\approx 4000$ $u$ is not particularly
small. On the other hand, it is sufficiently small for the
parametrizations of \cite{schloms} to be justified.

\section{Discussion}

We have seen that self-avoiding walks on the hypercubic lattice in four
dimensions show clear logarithmic corrections. These corrections are in
perfect agreement with the predictions of the renormalization group,
though they would be very poorly described by the asymptotically leading
approximation. Thus claims \cite{havlin,aragao,rens} that exactly these
leading terms have been seen in simulations must be viewed with some
reservation. On the other hand, our numerical results agree perfectly
with simulations reported in \cite{rapa}, although the conclusion drawn
in that paper seems to be wrong. Though our data are perfectly fitted
by power laws in $\ln N$ --- in particular,
\be
   R_N^2/N \sim [\ln N]^{0.31},    \label{wrong}
\ee
these do not represent the asymptotic behavior.

In order to understand the discrepancy with \cite{aragao,rens}, we
have repeated their computation of $\langle N_R\rangle_p$ and their
subsequent analysis, but with several modifications: \\
(i) using our algorithm, we calculated $\langle N_R\rangle_p$ not only
for a single fixed value of $R$ ($R=7$ in \cite{aragao,rens}), but
for all $R<\sqrt{500}$;\\
(ii) due to the efficiency of our algorithm (and improved hardware)
we have much higher statistics;\\
(iii) we use data much closer to the critical point: $p_c-p\geq 0.0012$
compared to $p_c-p\geq 0.018$; and \\
(iv) in the analysis we use our very accurate estimate for $p_c$.

Assuming that the decay of the two-point function is dominated by
the smallest mass $m$ in the model (the inverse correlation length
$\xi$), the authors of \cite{aragao,rens} obtain
\be
   \langle N_R\rangle_p \sim R \,{dm\over dp}    \label{ara}
\ee
with
\be
   m \equiv 1/\xi \sim \sqrt{p_c-p}\, [\ln (p_c-p)]^{-\cal N}\,,
            \quad {\cal N}=1/8\;.
\ee
Thus, $\langle N_R\rangle_p/R$ should be a function of $p$ alone,
independent of $R$. Our data shown in fig.10 clearly demonstrate
that this is true only for $R>>\xi$, while
\be
   \langle N_R\rangle_p \sim R\, \ln R
\ee
for $R<\xi$. This is not surprising: the decay of the correlation
function should be dominated by the smallest mass only for $R>>\xi$,
while effects of the effective coupling should be visible at shorter
distances. Indeed, keeping $R$ fixed while $p\to p_c$ we encounter
the region where the data should be analysed via a short distance
expansion, and eq.(\ref{ara}) is no longer justified. According to
fig.10, $R=7$ is not yet in the regime where eq.(\ref{ara}) holds for
the range of $p$ considered in \cite{aragao}. This gives already a
first indication that this coupling is larger than anticipated in
\cite{aragao} (we might add that the fluctuations seen in fig.10 are
not statistical, but are lattice effects).

In fig.11 we show our data for $R=7, \sqrt{150}$ and $\sqrt{500}$
together with those of \cite{aragao}, and with the mean-field prediction
\be
   {\langle N_R\rangle_p\over R}\; \propto \;{1\over \sqrt{p_c-p}}.
\ee
On the one hand we see that the slopes deviate strongly from mean field
prediction, much more than they differ between different values of $R$.
On the other hand, we see the dramatic effect of a wrong choice of
$p_c$. Indeed, while we use our value of $p_c$ when plotting our own
data, we show the data of \cite{aragao} twice: once plotted using our
$p_c$, and once using their own estimate of $p_c$. We see that the
latter reduces the deviation from mean field considerably. Thus we
see that the logarithmic corrections are indeed larger than estimated
in \cite{aragao} (in spite of the fact that the too small value of
$R$ alone would have led to their {\it over}estimation!), but the
systematic uncertainties and the lack of higher order predictions
from field theory prevent a more detailed analysis. The same
comments should hold for the data of \cite{rens}, as this author
used the same values of $R$ and of $p$ as \cite{aragao}, and
his estimated $p_c$ was even further from our estimate.

To what chain lengths would we have to go in order to see the
asymptotic behavior $R_N^2/N,\; C_N/\mu^N \sim [\ln N]^{1/4}$? Our
renormalization group calculation can easily be extended to larger $N$
and show that eq.(\ref{wrong}) no longer would give a good fit above
$N=10^4$, but even at $N=10^7$ the effective exponent is $\approx
0.285$ instead of $1/4$.

The analysis in the field theoretic framework stresses the importance
of a good quantitative form of the renormalization group mapping. For
the amplitudes then a low order approximation seems sufficient. This
supports previous findings, both in polymer physics~\cite{schaef}
 and in the physics of liquid helium~\cite{kraus}. If a good
form of the mapping were missing, this would justify the often
expressed pessimism about the possibility to see logarithmic
corrections. Such a situation prevails e.g. for $\theta$-polymers
in three dimensions. There only leading terms in the sense of
eq.(\ref{rr-dupl}),(\ref{cc-dupl}) have been
computed \cite{dupla}, and they disagree badly with simulations
\cite{theta}. Although the disagreement there is even worse than the
disagreement with the leading terms in the present case, one might
suspect that also there the poor representation of the renormalization
group mapping employed is at the root of the problem.

The present analysis showed large subleading corrections because of the
rather large value of the renormalized coupling constant, even for our
longest chains. This raises the question whether the asymptotic behavior
could be seen for much shorter chains in models with weaker repulsion
between chains. Two such models come immediately into mind: chains with
attraction between neighboring sites ($\theta$ polymers above $T_\theta$),
and the Domb-Joyce model \cite{domb-joyce}. In the latter, $n>1$ monomers
can occupy the same site, but the contribution to the partition function
gets a weight $(1-w)^{n-1}$ for each multiple occupancy. For sufficiently
small $w$ one is arbitrarily close to the free case, and also the
renormalized coupling constant should be small.

This work was supported by DFG, SFB 237, and by the Graduiertenkolleg
``Feld\-theoretische un numerische Methoden in der Elementarteilchen-
und Statistischen Physik".

\eject

\eject

\section*{Figure Captions:}

{\bf Fig.1:} Plot of the logarithm of the swelling ratio, $\ln(R_N^2/N)$,
against $\ln(\ln N)$. The solid line
are our data with $N$ taken as the number of bonds; the dotted line
represents these data with $N$ as the number of sites. The diamonds are
the data from \cite{rapa}, and the dashed line is the fit with
eq.(\ref{rr-dupl}). The dashed-dotted line indicates the slope predicted by
the leading term given in eq.(1). The statistical errors of our data
are roughly $\propto \sqrt{N}$. They are thus largest for $N=4000$, where
$\Delta \ln R_N^2 = 0.0005$.

{\bf Fig.2:} Semi-logarithmic plot of $C_N p^N$ against
$\ln N$. The plot shows our MC data for 3 different values of $p$
very near $p_c$, the fit with eq.(\ref{cc-dupl}), and the exact enumeration
data from \cite{mcd}. The statistical error of our data for $N=4000$ is
$\Delta C_N/C_N = 0.0013$.

{\bf Fig.3:} Plot of $(R_N^2/N)/(C_N p^N)$ against $1/\ln(10 N)$. According
to eq.(\ref{rc-supl}) a straight line with negative slope is expected for
$p=p_c$, with a slope as indicated by the dotted line.

{\bf Fig.4:} Function $\sigma_0(N)$ against $N$ (full line). Also shown is
$z_R^{(\alpha)}(N)$ as obtained by the Pad\'e approximant
eq.(\ref{zpade}) (dashed line).

{\bf Fig.5:} Function $\sigma_1(N)$ against $N$, for the same 3 values of $p$
as in fig.2.

{\bf Fig.6:} Functions $z_R^{(\alpha)}(N)$ (full line) and $z_R^{(C)}$
(dashed). The latter is again shown for the  same 3 values of $p$.

{\bf Fig.7:} Panel (a): function $W[z_R^{\alpha}]$ (full line) and the
``derivative" of $z_R^{\alpha}$ with respect to $\ln N$ (dashed line).
Both curves should agree up to higher orders in $z_R$. Panel (b) shows the
same data, but in the definition of $\alpha_N$ we have replaced $N$ by
$N+0.17$.

{\bf Fig.8:} Fit to the swelling ratio $\alpha_N$ with $\l_0=1.0436$ and
$n_0=0.275$.

{\bf Fig.9:} (a) Fit to $C_N/\mu^N$ with $c_0=0.870$ and $n_0=0.316$; (b)
fit to $\alpha_N$ with the same $n_0=0.316$, with $\l_0=1.0466$, and with
$n_1=0.56$.

{\bf Fig.10:} Average chain lengths at fixed $p$ against their
end-to-end distance $R$. Actually, the ratio $\langle N_R\rangle_p/
R$ is plotted. The values of $\ln(1/p)$ (from top to bottom) are:
1.9143, 1.915, 1.916, 1.918, 1.921, 1.925, 1.932, 1.948, 1.966, 1.984,
2.002, 2.021, 2.066,2.100, and 2.150.

{\bf Fig.11:} Full lines: $\langle N_R\rangle_p/R$ against $p_c-p$
for $R=7, 12.25, 22.36$ (from bottom to top); dotted line: slope
$\langle N_R\rangle_p\propto (p_c-p)^{-1/2}$ predicted by mean field
theory; diamonds: data of \cite{aragao} (for $R=7$), using our value
of $p_c$; crosses data of \cite{aragao}, using their value of $p_c$.

\end{document}